\documentclass[10pt,sigconf,prologue,dvipsames,numbers,sort&compress]{acmart}

\renewcommand\footnotetextcopyrightpermission[1]{} % removes footnote with conference information in first column
\newcommand{\figspace}{\vspace{-3mm}}

\usepackage{comment}
\usepackage{enumerate}
\usepackage{color}
\usepackage{relsize}
\usepackage{subfigure}
\usepackage{tabularx}
\usepackage{booktabs}
\usepackage{listings}
\usepackage{amssymb}
\usepackage{wrapfig}
\usepackage{algorithm}
\usepackage{algpseudocode}
\usepackage[utf8]{inputenc}
\usepackage[english]{babel}
\usepackage{amsthm}
\usepackage{mathtools}
\usepackage{lipsum} % Used for inserting dummy 'Lorem ipsum' text into the template

\begin{document}

\title{SciChain: Trustworthy Scientific Data Provenance }

\author{Abdullah Al-Mamun$^\dagger$, Dongfang Zhao$^{\dagger\S}$ 
\\
    \textit{
       % \footnotesize{
            $^\dagger$University of Nevada, Reno \hspace{3mm}
            $^\S$University of California, Davis 
        % }
    }
}

\begin{abstract}
The state-of-the-art for auditing and reproducing scientific applications on high-performance computing (HPC) systems is through a data provenance subsystem. 
While recent advances in data provenance lie in reducing the performance overhead and improving the user's query flexibility,
the fidelity of data provenance is often overlooked:
there is no such a way to ensure that the provenance data itself has not been fabricated or falsified.
This paper advocates to leverage blockchains to deliver immutable and autonomous data provenance services such that scientific data are trustworthy.
The challenges for adopting blockchains to HPC include designing a new blockchain architecture compatible with the HPC platforms and, more importantly, a set of new consensus protocols for scientific applications atop blockchains.
To this end, we have designed the proof-of-scalable-traceability (POST) protocol and implemented it in a blockchain prototype, namely SciChain, the very first blockchain system for HPC.
We evaluated SciChain by comparing it with multiple state-of-the-art systems; 
Experimental results showed that SciChain guaranteed trustworthy data while incurring orders of magnitude lower overhead.
\end{abstract}

\maketitle

    \figspace
\section{Introduction}

\subsection{Motivation}

The \textit{de facto} way to audit and reproduce scientific research and data is through data provenance,
which tracks the entire lifespan of the data during the experiments and simulation at various phases such as data creation, data changes, and data archival.
Data provenance plays a critical role in guaranteeing the validity of scientific discoveries and research results,
as data fabrication and falsification could happen to meet research objectives or personal interests or both.
For instance, the National Cancer Institute found 0.25\% of trial data are fraudulent in the year of 2015~\cite{datafraud_clinic15}.
In earth sciences, scientists emphasized the importance of maintaining data provenance in achieving the transparency of scientific discoveries~\cite{provenance_earthscience15}.

Conventional provenance systems can be categorized into two types:
centralized provenance systems and distributed provenance systems.
One popular centralized provenance system is SPADE~\cite{spade_2012},
where the provenance (from various data sources) is collected and managed by a centralized relational database.
Domain-specific systems based on such centralized design paradigms are also available in biomedical engineering~\cite{tclark_jbs14}, computational chemistry~\cite{pettersen_jcc04}, among others.
Although having been reasonably adopted by various disciples,
the centralized provenance systems are being increasingly criticized due to the exponentially-grown data:
the centralized provenance system becomes a performance bottleneck and a single point of failure, 
and to this end,
we witness the boom of various distributed approaches toward \textbf{scalable provenance}~\cite{dzhao_cluster13,ddai_pact17}.
Indeed, those distributed provenance systems,
mostly built upon distributed file systems as opposed to centralized databases,
eliminated the performance bottleneck and proved to deliver orders of magnitude higher performance than centralized approaches.

As a double-edged sword, however, distributed provenance systems pose a new concern~\cite{provenance-security-bates2015trustworthy} on the provenance itself:
\textbf{while the provenance is supposed to audit the execution of the application, 
who then should audit the provenance?}
Do we need to build the provenance of provenance?
So the recursion goes on and on, indefinitely.
Note that, this concern was not that critical in a centralized approach as long as we can, which is the case, apply robust reliability mechanisms to the centralized node,
however, it becomes a very challenging problem for all the participating nodes in a (large-scale) distributed system:
if any single node of the entire deployment is compromised, 
the provenance as a whole becomes invalid.
To this end, \textbf{decentralized provenance systems} were recently proposed inspired by blockchains. %\abdullah{Added lineage chain}
These systems (e.g., ProvChain~\cite{prov-xliang_ccgrid17}, SmartProvenance~\cite{aram_codaspy18}, LineageChain~\cite{provenance-blockchain-ruan2019fine}) are also called blockchain-based provenance systems that are both temper-evident and autonomous, thus guarantee trustworthy data.
The key idea of blockchain-based provenance systems is:
\textit{instead of storing the data on a single node or split it into $n$ nodes,
let us \textbf{replicate} the data and maintain a \textbf{hashed linkedlist} for each copy of the data}.
The \textit{replication} guarantees the provenance is tolerant to a certain degree of fault (e.g., $\lfloor \frac{n-1}{3} \rfloor$ in a Byzantine system),
and the \textit{hashed linkedlist} guarantees that the provenance data cannot be tampered with without being noticed by a simple hash verification.

Nonetheless, multiple issues must be addressed before blockchain-based provenance systems come to practical usage.
For instance, there is a series of concerns on resource utilization: the space efficiency is low, the network bandwidth consumption is high, the CPU cycles are ``wasted'' for meaningless mining, to name a few.
Besides, all these blockchain-based provenance systems are built in such a way that the underlying blockchain infrastructure is a black box, and the provenance service works as a higher-level application by calling the programming interfaces provided by the blockchain infrastructure such as Hyperledger Fabric~\cite{hyperledger} and Ethereum~\cite{ethereum}.
In the best case, the provenance services might miss optimization and customization opportunities because the former cannot modify the lower blockchain layer;
to make it worse, the applicability of those blockchain-based provenance systems is constrained by the underlying blockchain infrastructure.

In the following, we highlight some limitations exhibited from existing blockchain-based provenance systems in the context of scientific computing and high-performance computing (HPC) systems.
(1) If the compute nodes of the scientific experiments have no local disks,
    which is very common for large-scale HPC systems, 
    then the blockchain-based provenance system (at least in its current form) becomes useless as all blockchain systems require node-local persistent storage. 
(2) If the consensus protocol of the underlying blockchain system is inappropriate for the scientific applications, then again, the provenance system becomes useless. To make the matter more concrete, the popular proof-of-work (PoW) consensus protocol is taken by many public blockchains that can hardly be applied to large-scale scientific simulations: it is nonsense to ask every single compute node to (re-)run the experiments.
(3) If the experiment testbed has a multi-tiered storage architecture (e.g., burst buffers, I/O nodes, remote parallel file systems), the blockchains will be agnostic of such heterogeneity, 
    leading to sub-optimal performance.

In summary, a highly desired provenance system for scientific applications should be crafted with a balance between scalability, reliability, and applicability.
Unfortunately, existing provenance systems failed to meet the above unique requirements from scientific computing the HPC communities.
Of note, a recent work~\cite{aalmamun_bigdata18} indeed proposed a blockchain-like provenance system deployed to diskless compute nodes (the so-called \textit{in-memory blockchain}, IMB),
IMB simply took the remote parallel filesystem as a pseudo independent local disk,
which brought limited insight and was not evaluated against real-world blockchain systems.

\subsection{Proposed Approach}

This paper proposes a new decentralized approach to manage the data provenance of scientific applications deployed to HPC systems. 
Rather than only taking an existing blockchain system as a block box,
we hack into blockchain internals to improve the applicability and performance of provenance services built upon blockchains.
Specifically, we design a new blockchain architecture supporting multi-tier storage and then devise new consensus protocols aiming to optimize the decentralized provenance services in an HPC environment. 

Specifically, this paper makes the following contributions:
\begin{itemize}
\item We propose a new architecture for secure and reliable decentralized data provenance on HPC systems.
The new architecture is tailored to the HPC environment: compute nodes can maintain the blockchain in local memory while using distributed shared ledger as a persistent medium for enhanced reliability and as a precaution for any catastrophes (e.g., compute nodes failure or restart). 

\item We design a set of consensus protocols, namely, proof-of-scalable-traceability (POST), for validating applications' data provenance following a \textit{push/pull} mechanism that promises memory optimization.
The key idea of POST is that the consensus comes not only from the fellow compute nodes through PoW but also from the remote shared storage through proof-of-extended-traceability (POET). 
POET comes into action only if the compute nodes are unable to reach consensus,
thus significantly reducing the communication overhead between the compute nodes and the shared storage.

    \item We prove the liveness of the proposed consensus protocols.
In particular, we show that all of the participating nodes would eventually reach consensus in front of arbitrary failures.

    \item We implement a system prototype, SciChain, and experimentally verify the system's effectiveness with more than one million transactions derived from both micro-benchmarks and real-world applications on up to 1,024 nodes.
\end{itemize}

The remainder of this paper is organized as follows. 
We review the background and related work on blockchains and data provenance in~\S\ref{sec:background}.
\S\ref{sec:protocol} presents the POST protocol along with its complexity analysis and liveness proof.
We detail the implementation of SciChain, a fully-fledged blockchain system with the POST protocol in~\S\ref{sec:impl}.
\S\ref{sec:eval} reports experimental results by comparing SciChain with other state-of-the-art systems.
We finally conclude the paper with future directions in~\S\ref{sec:conc}.

\section{Background and Related Work}
\label{sec:background}
A blockchain system is a distributed ledger that consists of multiple nodes that are either fully or partially trustworthy. Even while most of the nodes are honest, some nodes tend to exhibit Byzantine behavior because of the unexpected attack. All the nodes are responsible for generating, validating, and appending blocks with transactions in their local ledger as well as maintain a shared replica of a blockchain or ledger (i.e., a set of shared blocks of transactions) and global states. In a blockchain-like distributed ledger, more than 50\% nodes need to provide consensus to validate a block.

Two of the most popular types of blockchain systems so far,
in terms of access permissions,
are public and private blockchains. 
Other types are derived from these two basic types, such as inter-blockchains~\cite{inter,dzhao_cidr20}.
In the private blockchain system (i.e., permissioned network), a node requires special permission. The network initiator then either verifies the node or prepares a set of rules to verify the node. This mechanism puts a restriction on who is allowed to participate in the network, and only in certain transactions. Once a node joins the network, it starts to play a role in maintaining the blockchain in a decentralized manner. In a public blockchain network, anyone can join and participate. The network follows an incentivizing mechanism to encourage more participants to join the network. However, public blockchain systems have mainly two downsides. First, the openness of public blockchain supports weak security. Second, to fill up the security gap, each node requires to solve a puzzle and the difficulty in the puzzle needs to be increased along with the scaling of the node.

Several consensus protocols have been followed in the blockchain systems in order to validate blocks such as proof-of-work (PoW), proof-of-stake (PoS), and practical byzantine fault tolerance (PBFT). In PoW, each node in the network needs to solve a complex puzzle that needs a significant amount of computation, and the node that solves the puzzle is incentivized~\cite{tutorial-maiyya2018}. In PoS, the creator of a new block is chosen in a deterministic way, based on the wealth (i.e., stake) of the node and there is no reward for mining a block~\cite{Kiayias_2017}. In PBFT, all of the nodes (partially trusted) within the system extensively communicate with each other in order to reach an agreement about the state of the system through a majority~\cite{crypto-algorand-Gilad:2017}. 

\begin{figure}[!t]
%\abdullah{Change this}
    \centering
    \includegraphics[width=65mm]{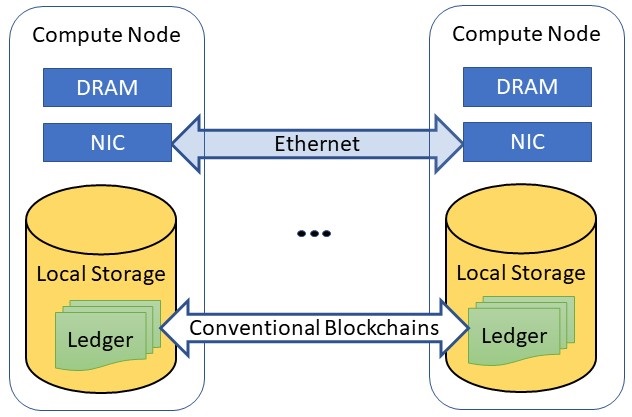}
    \caption{Conventional blockchain architecture.
    }

    \label{fig:blockchain}
        \figspace
\end{figure}

\begin{figure}[!t]
    \centering
    \includegraphics[width=65mm]{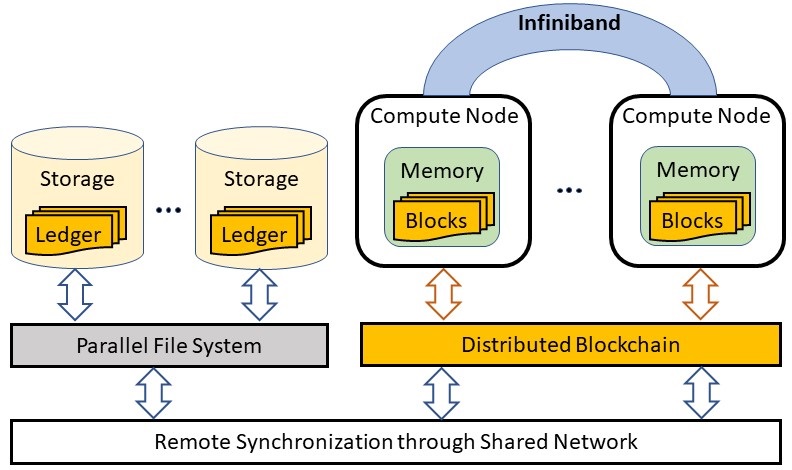}
    \caption{HPC blockchain architecture.
    }
    %\abdullah{Done: Will revise the figure to highlight the push/pull}
    \label{fig:dml}
    \figspace
\end{figure}

Figure~\ref{fig:blockchain} shows the system architecture of conventional blockchains deployed to a shared-nothing cluster.
Regardless of private (Hyperledger~\cite{hyperledger}) or public (Ethereum ~\cite{ethereum}),
all existing blockchain systems assume that the underlying computer infrastructure is shared-nothing:
the memory subsystems and I/O subsystems are all independent on the participant nodes who are often connected through commodity networking such as Ethernet.
In contrast, Figure~\ref{fig:dml} illustrates one possible blockchain deployment on HPC.

Data provenance tracks the source of data and the movement among the different data sources~\cite{provenance-track-wu2019provcite}. It facilitates the debugging process that helps to ensure the reproducibility of results. Therefore, security and privacy issues, e.g., integrity, confidentiality, and availability, are major concerns~\cite{provenance-security-bates2015trustworthy}. %\abdullah{need to add more generic provenance} in data provenance. 
State-of-the-art filesystem-based provenance systems~\cite{ddai_pact17,dzhao_cluster13},
and their variants such as in-memory blockchains~\cite{aalmamun_bigdata18},
lack abstractions that are essential for enabling trustworthy and reliability in provenance, 
calling for evaluations of the practical effectiveness of trustworthy provenance techniques~\cite{prov-davidson2017}. 

\section{Protocols}

\label{sec:protocol}

\begin{figure}[!t]
    \centering
    
    \includegraphics[width=65mm]{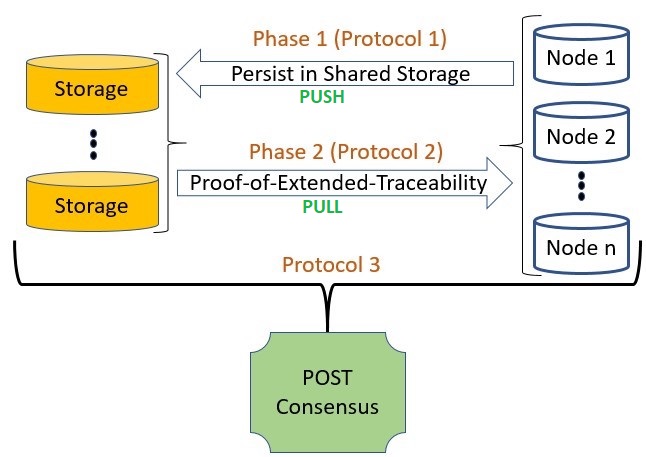}
    \caption{POST consensus.
    }
    \label{fig:POST-consensus}
    \figspace
\end{figure}

The proposed proof-of-scalable-traceability (POST) protocol, 
which is designed to optimize memory consumption, 
consists of two phases of protocols, as shown in Figure \ref{fig:POST-consensus}. 
In the first protocol (i.e., Phase 1), \textit{persist-in-shared-storage} (i.e., \textit{Push method}), the shared storage is being leveraged in order to keep ledger replica with minimum communication overhead as all ledgers on compute nodes are necessarily volatile.
The second protocol (i.e., Phase 2), \textit{proof-of-extended-traceability}, further extends the first protocol by pushing the storage node into a more active position:
whenever a new block is created,
it will be validated by the compute nodes first, and if more than 50\% compute nodes are compromised, the distributed shared storage participates in the validation process (i.e., \textit{Pull method}).
Details of each protocol are as follows.

\subsection{PSS: Persist in Shared Storage}
The protocol is described in Protocol~\ref{alg:persistent}.
The key idea is, a batch of new transactions are stored to the blockchain and persisted to a parallel file system (e.g., GPFS),
as shown in Line 5 and Line 10. That is, the blocks of transactions are appended to the compute nodes' memory first based on the validation achieved from Protocol~\ref{alg:mine} shown in Line 4. %\abdullah{push/pull is highlighted here} 
It should be noted that we only keep the most recent blocks within the scope of the compute nodes' memory to minimize the memory overhead. This is because, if more than 50\% compute nodes can not validate a new block based on the current blocks available in memory, which is rare, they can always use the \textit{Pull method} (i.e., Protocol~\ref{alg:mine}) to retrieve the latest blocks from the shared storage to synchronize. If more than 50\% compute nodes can validate the block as shown in Line 9, the block is persisted in the shared storage as well (i.e., \textit{Push method}). 
Doing so adds an extra layer of reliability to the data on volatile memory. However, the block validation and replication process both in the compute nodes and in the shared storage should not exhibit significant overhead. This is addressed in Line 2, where we check first the shared storage before starting the block appending process. 
That is, we start accessing the compute nodes only when the block is not validated and appended already in the persistent medium (i.e., shared storage). Then the questions arise, what if the shared storage gets compromised, and how frequently the storage is accessed. It should be noted that the persistent medium is accessed only once when more than 50\% compute nodes agreed on the validation of a block, as shown in Line 9, which ensures exceptionally light communication among the compute nodes and shared storage while making the shared storage secure.

\begin{algorithm}
\floatname{algorithm}{Protocol}
    \caption{Persisting in-memory blocks to shared storage} 
    \label{alg:persistent}
    \begin{algorithmic}[1]
        \Require 
        Compute nodes $M$ where the $i$-th node is $M^i$; 
        $M^i_B$ the local blockchain on $M^i$;
        a newly mined block $b$; 
        shared storage $S$;
        $S_B$ the blockchain copy on the shared storage;
        \Ensure Persist $b$ to $S$ and $M$ after validating it by 51\% compute node list $VNodes$ 
        \Function{Persist-in-Shared-Storage}{$b$, $M$, $S$}
        \If {$b \not\in S_B$} \Comment{Need only one look-up}
        
            \For {$M_i \in M$}
            \If {$b$ is valid with $M_B^i$} \Comment{Protocol ~\ref{alg:mine} }
               
                \State $M^i_B \gets M^i_B \cup b$ 
                \State $VNodes  \gets VNodes \cup M_B^i$
                
            \EndIf
            \EndFor
            \If {$|VNodes| > \frac{N}{2}$} \Comment{Protocol ~\ref{alg:consensus} }
                \State $S_B \gets S_B \cup b$ \Comment{Push method}
            \EndIf
          
        \EndIf
        \EndFunction
    \end{algorithmic}
\end{algorithm}

The theoretical time complexity of this protocol is $O(|M|)$, and $|M|$ could be a reasonably large number (e.g., tens of thousands of cores in leading-class supercomputers~\cite{cori}), which is on par with the original PoW consensus. 
We will demonstrate the effectiveness of the protocol in the evaluation section.
The correctness of Protocol~\ref{alg:persistent} is evident because the only change to the original PoW consensus is the data persistence,
which has nothing to do with the agreement between the compute nodes. The main goal of this first protocol is to help the consensus protocol (i.e., Protocol~\ref{alg:consensus}) in continuing the block validation process with the help of our proof-of-extended-traceability protocol (i.e., protocol~\ref{alg:mine}). 

\subsection{POET: Proof of Extended Traceability}
POET (i.e., Protocol~\ref{alg:mine}) works in two steps. First, it checks whether it can validate a block with the help of in-memory blockchains stored in the compute nodes, as shown in Line 3. If the in-memory blockchains are sufficient enough to validate a block, the block is then stored both in the compute nodes (i.e., Line 5) as well as in the shared storage (i.e., Line 20). However, if more than 50\% compute nodes fail, which is rare, the storage node then comes forward (i.e., \textit{Pull method}) to proceed further with the block validation process (i.e., Line 11). If the storage node can complete the validation process successfully, the block is then stored both in the compute node (i.e., Line 15) and the shared storage node (i.e., Line 12). It should be noted that the shared storage is accessed only once either when more than 50\% compute nodes agree on the validation of the block as shown in Line 20 or when the block is validated for the first time against the shared storage in case of more than 50\% compute nodes failure as shown in Line 12.

The benefits of this new validation method are two folds:
(i) A faster validation process achieved by the agreement of the compute nodes, thanks to the in-memory support, 
(ii) An extended and stable validation support is achieved from the shared storage that serves as a reliable persistent medium if more than 50\% compute nodes are compromised or fail to provide consensus. 
The time complexity of Protocol~\ref{alg:mine} is $O(|M|)$,
which is on par with the original PoW consensus.
However, %as mentioned above, i.e., benefit (iii),
it is possible to reduce the number of iterations if we assign more weight to the shared storage.

\begin{algorithm}
\floatname{algorithm}{Protocol}
    \caption{Proof-of-Extended-Traceability on storage} %(called by Protocol~\ref{alg:consensus})}
    \label{alg:mine}
    \begin{algorithmic}[1]
        \Require 
        Compute nodes $M$ where the $i$-th node is $M^i$; 
        $M^i_B$ the local blockchain on $M^i$;
        a newly mined block $b$; 
        shared storage $S$;
        $S_B$ the blockchain copy on the storage;
        %$H$ the hash list of blocks on the storage;
        \Ensure $b$ is validated by both the local SciChain ledger $M^i_B$ and the remote persistent ledger $S_B$, and then appended to all $M_B$'s and $S_B$.
        
        \Function{Proof-of-Extended-Traceability} {$b$, $M$, $S$}
        \For {$M_i \in M$}
            \If {$b$ is valid with $M_B^i$ }
               
                \If {$b \not\in M_B^i$} 
                      \State $M_B^i \gets M^i_B \cup b$ \Comment{Protocol \ref{alg:persistent}}
                      \State $VNodes  \gets VNodes \cup M_B^i$
                \EndIf
                %\If {$b \not\in H$}\Comment{one lookup is needed} 
                %       \State $S_B \gets S_B \cup b$ \Comment{Protocol \ref{alg:persistent}}
                %\EndIf
                \EndIf
        \EndFor
            \If {$|VNodes| <= \frac{N}{2}$} \Comment{Compute nodes fail?}
        %     \Comment{Protocol \ref{alg:consensus}}
             
               \If {$b$ is valid with $S_B$ and $b \not\in S_B$}\Comment{Pull method}
                 \State $S_B \gets S_B \cup b$ \Comment{Protocol \ref{alg:persistent}}
                 \For {$M_i \in M$} 
                 \If {$b \not\in M_B^i$}
                    \State $M_B^i \gets M^i_B \cup b$ %\Comment{Protocol \ref{alg:persistent}}
                 \EndIf
                 \EndFor
              \EndIf
            \Else \Comment{Persist valid block in storage}
              \State $S_B \gets S_B \cup b$ \Comment{Protocol \ref{alg:persistent}}
            \EndIf

       \EndFunction
    \end{algorithmic}
\end{algorithm}

\subsection{POST: Proof of Scalable Traceability}
\label{sec:ext-validator}

To achieve the consensus in POST as shown on Figure \ref{fig:POST-consensus}, both Protocol~\ref{alg:persistent} and Protocol~\ref{alg:mine} are included to achieve Protocol~\ref{alg:consensus},
by which it is ensured that most of the compute nodes (more than 50\%) ensure the validity of the newly proposed block's hash after solving the compute-intensive problem (i.e., the ``puzzle'') from all respective in-memory blocks previously-stored in all compute nodes and the shared storage. If 51\% compute nodes get compromised, it is guaranteed that at-least the shared storage can ensure the validity.
As shown in Line 3 of Protocol~\ref{alg:consensus}, 
all of the compute nodes in the network attempt to solve the puzzle for the newly proposed block, and the block will only be appended to the shared storage, and the compute nodes (shown at Lines 13 and 15) if more than 50\% nodes agree on the validity of the new block. If at-least 51\% compute nodes are unable to provide the consensus (i.e., Line 7), then the remote storage node starts to validate the block (i.e., Line 8) in order to achieve 51\% consensus jointly from compute nodes and remote storage node. This mechanism helps to reduce the communication overhead between the compute nodes and the remote shared storage during consensus reaching process, because, the shared storage will only be accessed if more than 50\% compute nodes fail for some reason (e.g., out-of-memory error or restart).

\begin{algorithm}
\floatname{algorithm}{Protocol}
    \caption{All compute and storage nodes reach consensus}
    \label{alg:consensus}
    \begin{algorithmic}[1]
        \Require 
        Compute nodes $M$ where the $i$-th node is $M^i$; 
        $M^i_B$ the local blockchain on $M^i$;
        a new block $b$; 
        shared storage $S$;
        $S_B$ the blockchain copy on the storage;
        \Ensure At least 50\% compute node list $VNodes$ who validate $b$ both with local blockchain and with remote persistent ledger $S_B$.

        \Function{POST-Consensus}{$b$, $M$, $S$}
        \For {$M_i \in M$} %\Comment{Protocol 1}
        \If {$b$ is valid with $M_B^i$} 
        \State $VNodes  \gets VNodes \cup M_B^i$ \Comment{Consensus from compute nodes}
        %\Else
        %\State continue \Comment{If a compute node fails continue to next}
        \EndIf
         \EndFor
         
         \If {$|VNodes| <= \frac{N}{2}$}\Comment{Checks  if 51\% nodes provide consensus}
                \If {$b$ is valid with $S_B$} 
                  \State $VNodes  \gets VNodes \cup S_B$ \Comment{Consensus from storage nodes}
            
              \EndIf

         \Else %{$| VNodes | > \frac{N}{2}$} 
              \Comment{Protocol \ref{alg:persistent}}
                 \For {$M_i \in M$}
                    \State $M_B^i \gets M^i_B \cup b$ 
                \EndFor
            \State $S_B \gets S_B \cup b$ 
            
        \EndIf
       
        \EndFunction
    \end{algorithmic}
\end{algorithm}

As the Protocol~\ref{alg:consensus} utilizes both Protocol~\ref{alg:persistent} and Protocol~\ref{alg:mine} and does impact the agreement formation between participant nodes and the storage node.
Therefore, we need to demonstrate that the new protocol indeed leads to consensus,
meaning that at least 50\% of the compute nodes are guaranteed to hold the longest (same) ledgers in front of arbitrary attacks. Therefore,
formally, we provide the following theorem.

\subsection{Liveness of POST}
\label{subsec:post_progress}
In this section, the non-blocking property of the POST protocol is demonstrated. That is, the consensus reaching process (i.e., voting) will not be blocked due to the fail/restart of the nodes. To be more explicit, there is no ``partial consensus'' (i.e., either \textit{reached} or \textit{cancelled}) as long as the failed nodes can eventually recover. In the context of transactions, it is also referred to as \textit{commit} and \textit{abort}. In distributed computing, this non-blocking property is also called \textit{liveness}.

As demonstrated in~\cite{stonebreaker_3pc83},
there are two necessary and sufficient conditions to ensure a non-blocking protocol:
\begin{enumerate}
    \item[C1] Between commit and abort, there exists no such state which could help us to make a decision. 
    \item[C2] There is no direct link between an indeterministic\footnote{An indeterministic state is defined as a state from which no final decision can be made.} state and a committed state.
\end{enumerate}

The proof states as follows.
To verify C1, let's assume the node $P$ who initiates the transaction fails after $(m-1)$ other nodes have validated $P$'s request.
That is, total $m$ nodes, including $P$ itself, have verified the transaction before $P$ are failed and restarted.
There are two cases to consider:
\begin{itemize}
    \item If $m <= \frac{N}{2}$, then $P$ has not committed anything because the execution is incomplete according to Line 7 of Protocol~\ref{alg:consensus}. 
    In other words, the state always indicates an \textit{abort},
    and the restarted $P$ node will simply forward the transaction to fellow nodes to start the verification.

    \item If $m > \frac{N}{2} + 1$, then $P$ should have persisted the change to the disk according to Line 11 of Protocol~\ref{alg:consensus} and Line 20 of Protocol~\ref{alg:mine}.
    In this case, when $P$ restarts, it will definitely lead to a \textit{commit} status.
\end{itemize}
Therefore, we see that no matter how many nodes have verified the transactions requested by $P$ before the latter fails, 
for each possible case, there is only one possible outcome.
That is, we never need to decide a commit or abort operation given a specific case.
C1 is thus satisfied.

It is simple to verify C2 because POST does not exhibit an indeterministic state. 
Once $P$ restarts, it simply checks whether the transaction is stored in the persistent storage.
If so, $P$ will mark the transaction completed---a ``commit'' state;
otherwise, $P$ will resend everything and restart the consensus procedure---an ``abort'' state. C2 is thus satisfied.

\section{Design and Implementation}
\label{sec:impl}

\subsection{Overview}

We have implemented a prototype system of the proposed blockchain architecture and consensus protocols with about 2,000 lines of Java code.
At this point, we only release the core modules of the prototype;
some complementary components and plug-ins are still being tested with more edge cases and will be released once they are stable enough. The source code is currently hosted on Github.%~\cite{scichain-github}.

\begin{figure}[!t]
    \centering
    
    \includegraphics[width=70mm]{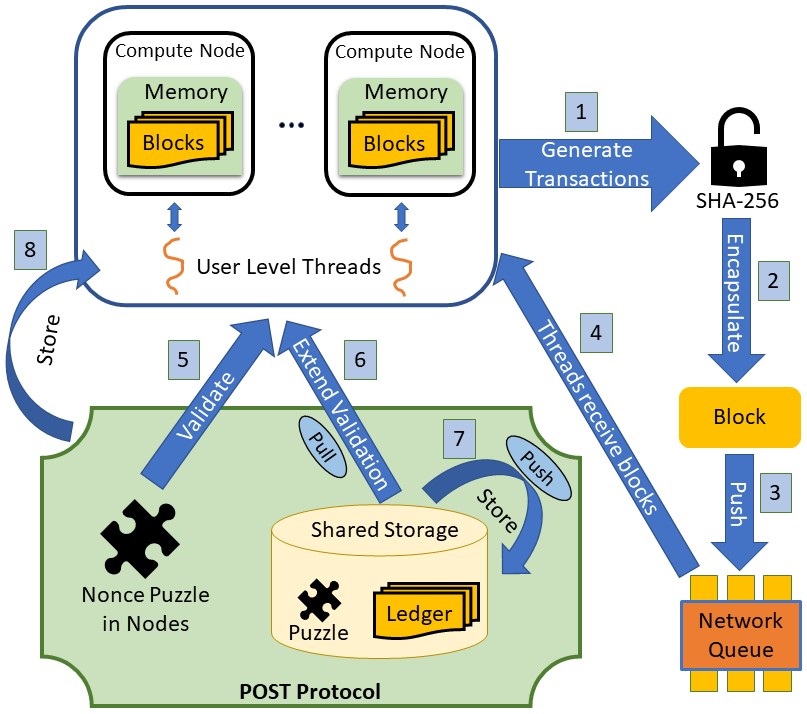}
    \caption{Overview of system interaction in SciChain.
    }
    %\abdullah{Done: I can also revise this figure to show push/pull mechanism}
    \label{fig:POST-interaction}
    \figspace
\end{figure}

Figure~\ref{fig:POST-interaction} illustrates the overview of the implemented architecture of the proposed SciChain. The prototype system currently runs on a virtualized environment where 
(i) each node is executed with a user-level thread, 
(ii) local ledgers are stored on as (distinct) files,
and 
(iii) network latency is throttled by a time delay parameterized by arbitrary statistical distributions (with average, variance, random seed).
The prototype is being packaged into dockers to be conveniently deployed to production systems.

As shown on Figure~\ref{fig:POST-interaction}, newly transactions generated by the nodes are first encrypted using SHA-256 algorithm~\cite{sha256} (step 1). 
Then, the transactions are encapsulated in a block by the respective nodes (step 2) before being pushed into a queue (step 3), 
from which the queued blocks are propagated across the network (step 4).
Finally, the blocks are validated both in compute nodes (step 5) and optionally, in the shared storage (step 6) (i.e., \textit{Pull}) to achieve POST consensus before storing them in the shared storage (step 7) (i.e., \textit{Push}) and the compute nodes (step 8). It should be noted that the shared storage takes part in the validation process (i.e., step 6), only if more than 50\% compute nodes are unable to provide consensus (i.e., step 5).

As proposed in~\cite{kzhang_icdcs18},
each blockchain system can be decomposed into four main components: 
ledger, consensus, cryptography, and smart contract. 
Because this paper focuses on ledger and consensus,
we take SHA-256~\cite{sha256} as the cryptography function for the chained blocks.
The current system implementation does not support smart contract:
we wrap up the applications using pseudo transactions with dummy numerical values such that the prototype can take in a variety of applications.
The remainder of this section focuses on the ledger and consensus implementation.

\subsection{Node Threads}
\label{sec:impl_node}
An unique user-level thread is generated for representing each node. 
All the properties of the node, 
such as node ID, a pointer to the local file, node type (storage or sensor), are stored in a \texttt{Node} object. 
More details about the data and storage structure will be discussed in~\S\ref{sec:impl_model}.

The nodes in our system prototype are implemented with user-level threads spawning transactions in random orders with a fixed time interval. 
The threads are responsible for packing the transactions into blocks;
each block comprises at least twelve transactions such that the launching overhead is amortized and negligible to each transaction.
The blocks are then pushed into the network queue that will be discussed in the following section~\S\ref{sec:impl_network}.
The nodes are also responsible for processing the block (e.g., mining block in Bitcoin),
which will be discussed in~\S\ref{sec:impl_POST}.

\begin{figure}[!t]
    \centering
    \includegraphics[width=85mm]{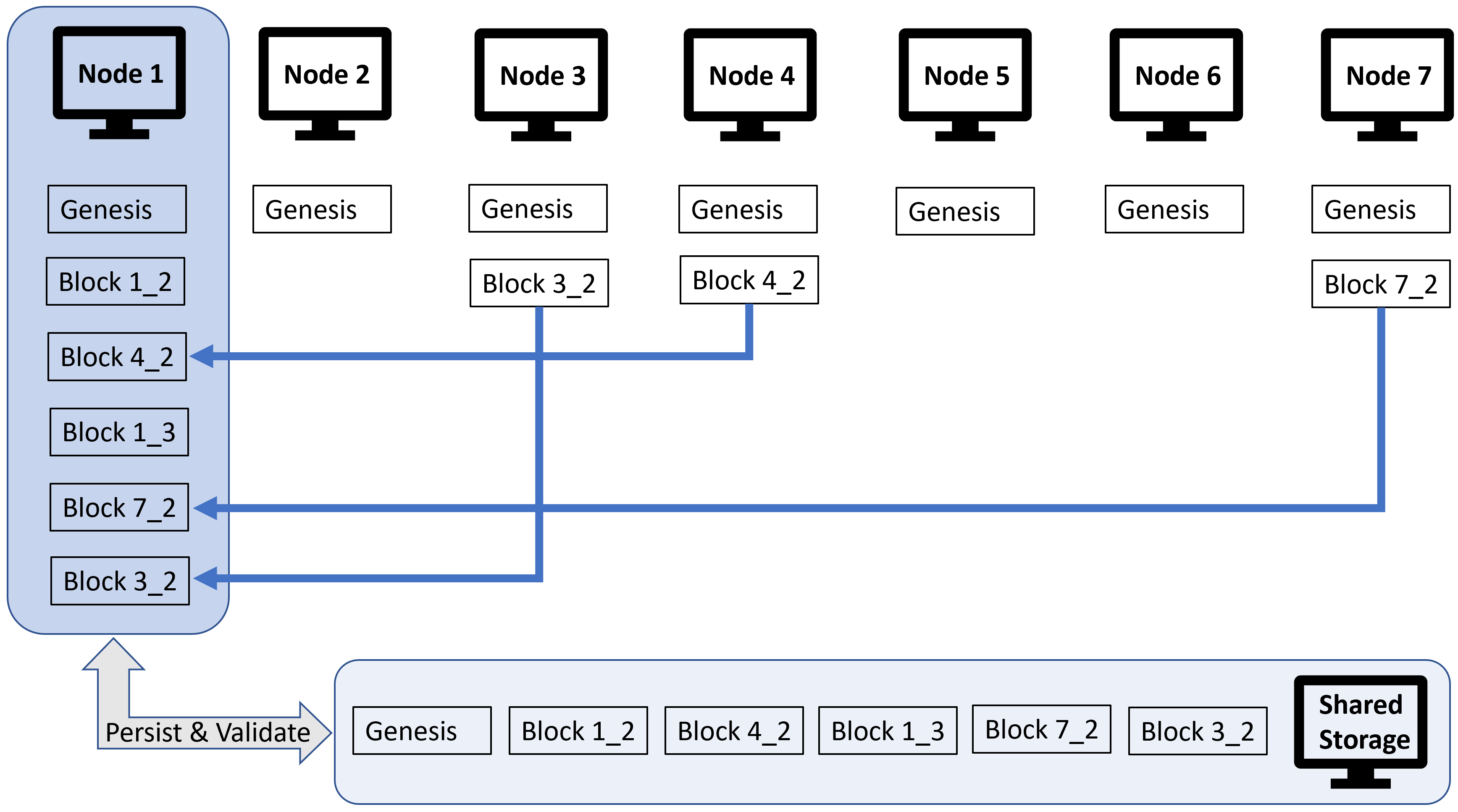}
    \caption{Data structure of SciChain transactions. 
    }
    \label{fig:ledger}
    \figspace
\end{figure}

\subsection{Network Queues}
\label{sec:impl_network}
All of newly submitted blocks by the nodes are pushed into a \texttt{FIFO queue}.
Similarly to how the blocks of transactions are spawned,
a block also periodically pops out from the head of the queue at a fixed time interval, which then is transferred to the connected nodes.

Because there are different latencies among the nodes in the real network based on different kinds of connection types; therefore, in the proposed solution, each peer connection is set with specific network latency. 
For instance, the InfiniBand connection latency between two nodes is 2 microseconds~\cite{rdma_latency}, whereas, in the case of Ethernet connection, the latency between two nodes is set with 250 microseconds~\cite{ethernet_latency}.

It could be argued that a queue might not deliver data at a sufficient rate to feed the network because a queue is a linear data structure that can hardly be parallelized for the sake of scalability.
This is alleviated by the following two approaches in our implementation.
First, we adjust the time interval more substantial than the latency for the queue to pop an element. 
In doing so, the overhead from the queue itself is masked completely.
Second, we implement the queue using a loosely-coupled linked-lists such that the queue can be split and reconstructed arbitrarily.

\subsection{Data Models and Storage}
\label{sec:impl_model}

The data structure for the proposed SciChain ledger is a linked list and stored in a row-wise table where each tuple corresponds a block, which references a list of transaction records stored in another table,
just like the traditional relational database tables. 
Figure~\ref{fig:ledger} shows a concrete example of the structure to store the blockchain on a specific node (i.e., Node 1).
The block \textit{genesis} indicates the very first block of the blockchain for that node. Each cell under the node represents a block. For example, under Node 1, Block 1\_2 indicates the second block created by Node 1 and is a child of the genesis block.
Block 1\_2 is also the parent of Block 4\_2, which is created by Node 4. 
Block 4\_2 is appended as the child of Block 1\_2 when Node 4 successfully appends this block and broadcasts it to other nodes. In addition to the inter-compute-node propagation, the new block is also persisted and validated on the \textit{Shared Storage} Node,
which is the bookkeeper of the ground truth of the decentralized blockchains.

\subsection{Consensus Protocols}
\label{sec:impl_POST}

The consensus protocol between compute nodes,
as a building block of the proposed POST consensus,
shares the same essence of the traditional PoW but simplifies the compute-intensive puzzle taken by Bitcoin~\cite{bitcoin}.
Specifically, our PoW does require a nonce number to mine and validate the new block.
Generally speaking, the nonce value serves as the knob controlling the balance between efficiency and trustworthiness.
The default nonce is set to one for the sake of efficiency, and we will report the sensitivity in~\S\ref{subsec:eval_sen}.

Firstly, all the newly created transactions generated in a node added to that node as well as are propagated to other peers %including the shared storage 
and eventually packed into a block by all the peers in the network for the preparation of starting the mining process.

Secondly, all the compute nodes will first attempt to validate and add the block to its local replica of the blockchain. The node who will first be able to solve the puzzle (i.e., mining) for the newly created block, will append the block in its local in-memory blockchain as well as in the remote storage and will receive the reward or credit. If more than 50\% compute nodes are unable to provide a consensus about the block, the remote storage will then come forward to validate the block through the POET protocol. % as it holds the highest stack (i.e., PoS).
If the block does not get validated in this round,
it is pushed into a waiting queue in the respective compute node so that the block can be processed by the node later.
This is the step where we follow the conventional proof-of-work (PoW) among the compute nodes but differently with a customized PoW (i.e., POST) especially for the scientific data provenance,
which is proof-of-extended-traceability by the ground truth held by the storage node in case of 50\% compute nodes failure.
 
Thirdly, when the block is appended %(i.e., after mining both by the node and optionally, if needed, also by the shared storage) 
successfully to the (local) blockchain by a node and the shared storage, it will then be propagated to the other peers in the network to make it validated; so that we can achieve the consensus.
In essence, the first node who actually is able to validate the block will be the node appending it (i.e., the new block) to the local blockchain and propagate the block to the entire network, 
and all other nodes in the network, 
after verifying the validity of the new block, will also update their local blockchains accordingly.  

\section{Evaluation}
\label{sec:eval}

\subsection{Experiment Setup}

\subsubsection{Testbed}

We leverage BlockLite~\cite{xwang_cloud19} for large-scale experiments.
The network latency of InfiniBand is two microseconds~\cite{rdma_latency};
The network latency of Ethernet is 250 microseconds~\cite{ethernet_latency}.
We deploy the system prototype mostly on 100 nodes except for the scalability test, where we use 1,024 nodes (Intel Core-i7 4.2 GHz CPU along with 32 GB 2400 MHz DDR4 memory).

\subsubsection{Evaluated Systems}

We evaluate the SciChain prototype against two other blockchain systems.
The first blockchain is a \textit{Conventional Blockchain} system deployed to a shared-nothing cluster with Ethernet connections.
The second blockchain is a \textit{Memory-only Blockchain} with high-performance networking interconnections (i.e., InfiniBand, RDMA) without any persistent storage;
this is not a practical solution due to the lack of data persistence but is considered as the performance upper bound of the proposed \textit{SciChain}.
We implement all three blockchain systems in Java and make a reasonable effort in optimizing all of them.

\subsubsection{Workload}

For micro-benchmarks,
the transaction format used in our evaluation is similar to funds transfer between bank accounts.
At the beginning of the transaction,
the system checks whether the submitted transaction is valid.
If so,
the statuses (balances) of two nodes are updated accordingly,
followed by the propagation of the updates to all other nodes in the network.
On average, each block contains about twelve transactions in our experiments.
We deploy more than one million transactions (1,036,303) to the system prototype,
and compare it to the other two baseline blockchain systems.

For real-world applications, 
The testing workload is derived from a trace of the FusionFS~\cite{dzhao_tpds16} filesystem initially deployed to a 1,024-node cluster at the Argonne National Laboratory.
The trace includes four real-world scientific applications:
Plasma Physics, Turbulence, AstroPhysics, and Parallel BLAST~\cite{pblast_2003}. 
We assign random numerical values to the I/O operations traced in FusionFS such that the SciChain prototype can take in the provenance in the same way as other blockchain systems.

\subsection{Trustworthy}

This section demonstrates that the proposed POST consensus can be achieved by more than 50\% of participants.
In other words, we want to show that introducing the shared storage as an additional node does not reduce the portion of good compute nodes to under 51\%.
To put it in another way, we want to verify whether the new consensus protocol leads to the same longest valid blockchain compared to other consensus protocols such as PoW from the compute nodes. 
Note that we prove the safety and liveness in Section~\ref{sec:protocol};
we will experimentally verify the trustworthiness and reliability here.

\begin{figure}[!t]
    \centering
    \includegraphics[width=65mm]{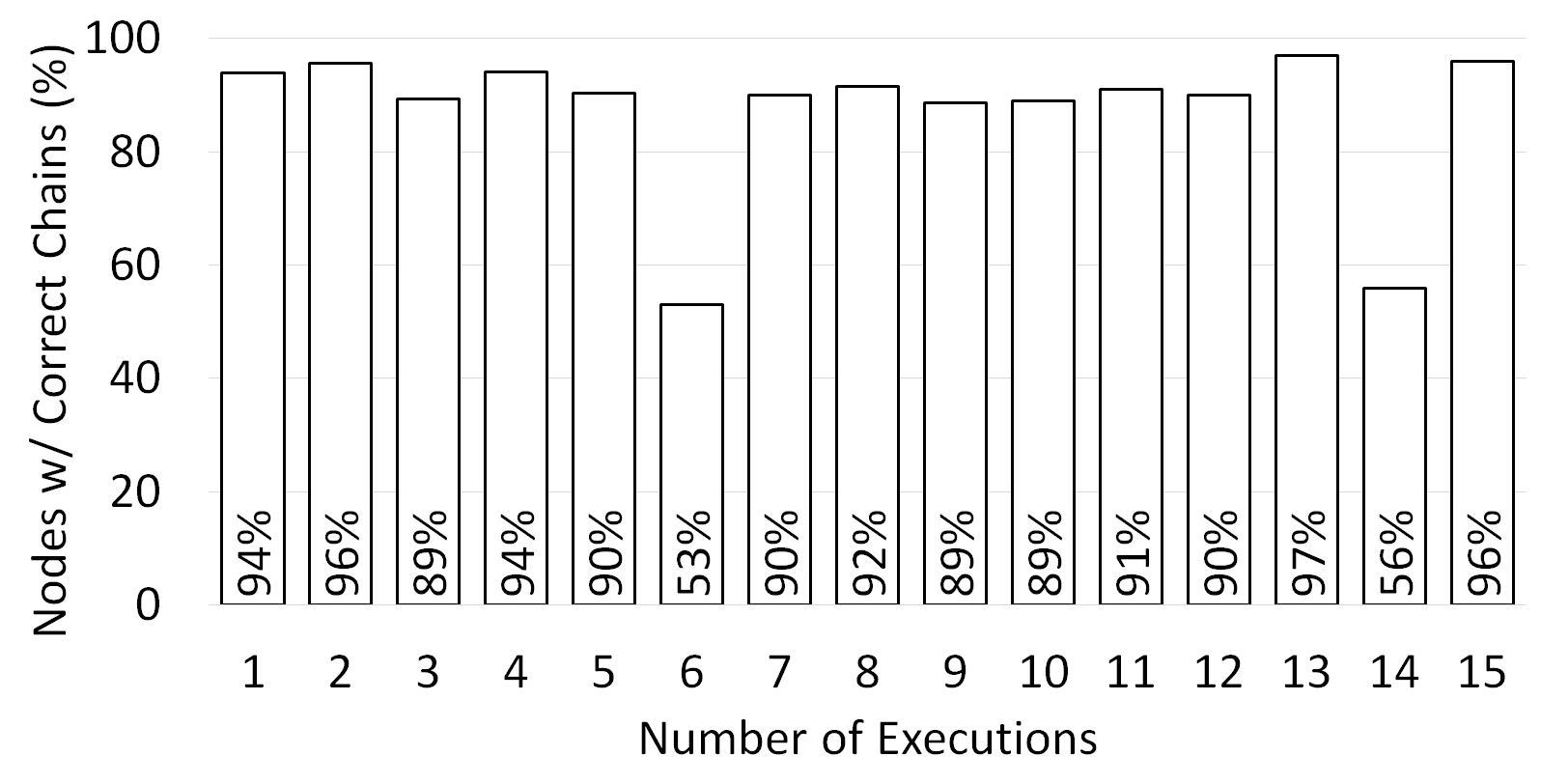}
     \caption{POST guarantees more than 50\% of nodes holding valid blocks.}
    \label{fig:post_reliability}
        \figspace
\end{figure}

To demonstrate the trustworthiness and reliability, we run the system prototype given random transactions for 10 minutes, and repeat the execution 15 times.
As we can see in Figure~\ref{fig:post_reliability}, all of the 15 executions lead to more than $50\%$ nodes holding the correct blockchains:
13 out of 15 executions yield more than 90\% validity while two executions exhibit lower ratios because we terminate the execution (i.e., 10 minutes) once more than 50\% nodes hold the correct blocks.
The bottom line is that we need to guarantee at least 51\% nodes' data are not tampered with, which is the case.

\subsection{Overhead}
This section reports the provenance overhead incurred by the proposed SciChain.
The memory-only, baseline system persists the data provenance to the disk with no security or audibility guarantees. 
We turn on SciChain atop the baseline and measure the end-to-end block processing time compared to the baseline performance.

\begin{figure}[!t]
    \centering
    \includegraphics[width=65mm]{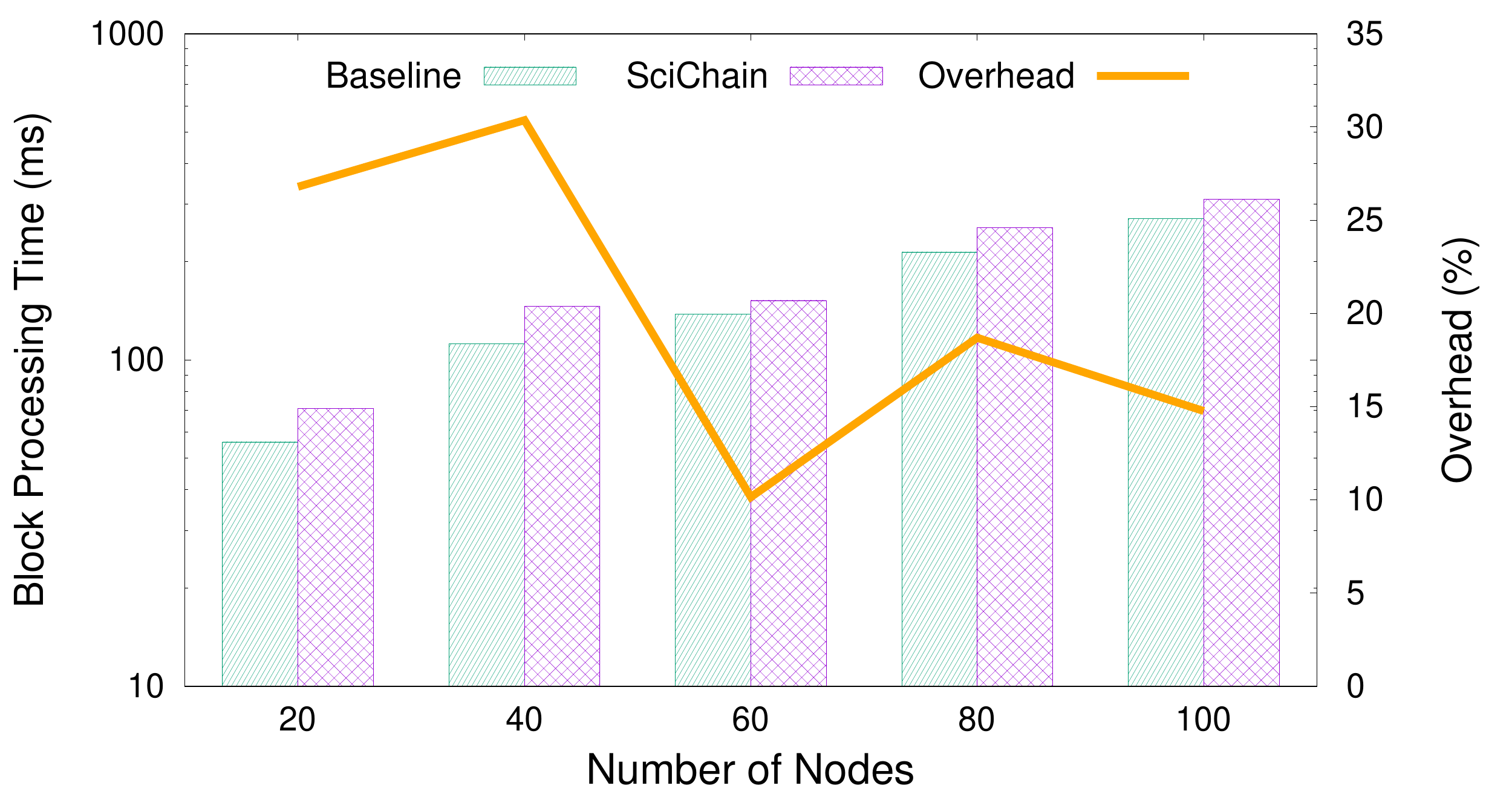}
     \caption{Latency overhead of SciChain compared to the Baseline. (Nonce = 1)}
    \label{fig:overhead_baseline}
\end{figure}

As shown in Figure~\ref{fig:overhead_baseline}, 
we observe the overhead incurred by the proposed SciChain is noticeable (25\% -- 30\%) at small scales of 20 and 40 nodes.
This is the price we have to pay to achieve high security.
The good news is, however, the overhead is reduced to under 20\% on larger scales;
in particular, 
the overhead is only 15\% on 100 nodes.
That is, unlike conventional provenance systems whose overhead increases proportionally to the number of nodes,
SciChain's overhead ratio does not significantly increase.

\subsection{Sensitivity}
\label{subsec:eval_sen}

\begin{figure}[!t]
    \centering
    \includegraphics[width=70mm]{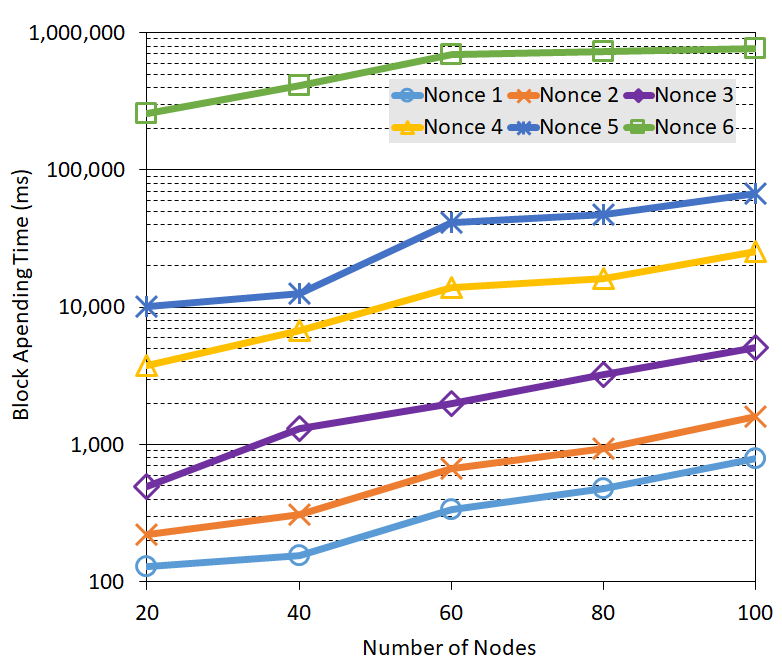}
    \caption{Performance sensitivity with different levels of puzzle difficulty.
    }
    \label{fig:nonce}
    \figspace
\end{figure}

A critical parameter in PoW-based blockchains is how to control the computational complexity to select the winner who can append the new block.
A large nonce implies high trustworthy and low performance,
while a small nonce indicates high performance and more vulnerability.
The canonical approach is to ask the node to find a ``nonce'' number,
combined with the hash value of the previous block,
which will result in a hash value of the current block in a specific format,
usually with a certain number of zeros.
We will use the short term \textit{nonce-number} to indicate the number of leading zeros required by the PoW consensus.
Because the proposed POST consensus is a variant of PoW, we want to study how the nonce number impacts the performance of the system.

Figure~\ref{fig:nonce} reports how different puzzle difficulties introduce the overhead;
we use `nonce-\textit{x}' to indicate that $x$ leading zeros are required.
In this experiment, $x$ ranges from one to six.
We see the trends of all nonce numbers follow a similar ascent pattern at different scales,
which is understandable because of the higher communication cost at larger scales.
A more interesting observation is that the performance impact from larger nonce values is more significant than the scale.
For instance, at nonce-5, the appending time is still under 70 seconds on 100 nodes;
In contrast, the appending time of nonce-6 on 20 nodes exceeds 200 seconds.
This result is essential for future HPC-crafted blockchains:
the scalability challenge will still exist, but the solution should be co-designed with a wise choice of nonce number.
One possible solution is to design an adaptive nonce value corresponding to the dynamic scales of applications,
however, it is beyond the scope of this paper, and we will leave this as an open question to the community and probably target this in our future work.

\subsection{Scalability}
\label{sec:eval_scale}

\begin{figure}[!t]
    \centering
    \includegraphics[width=75mm]{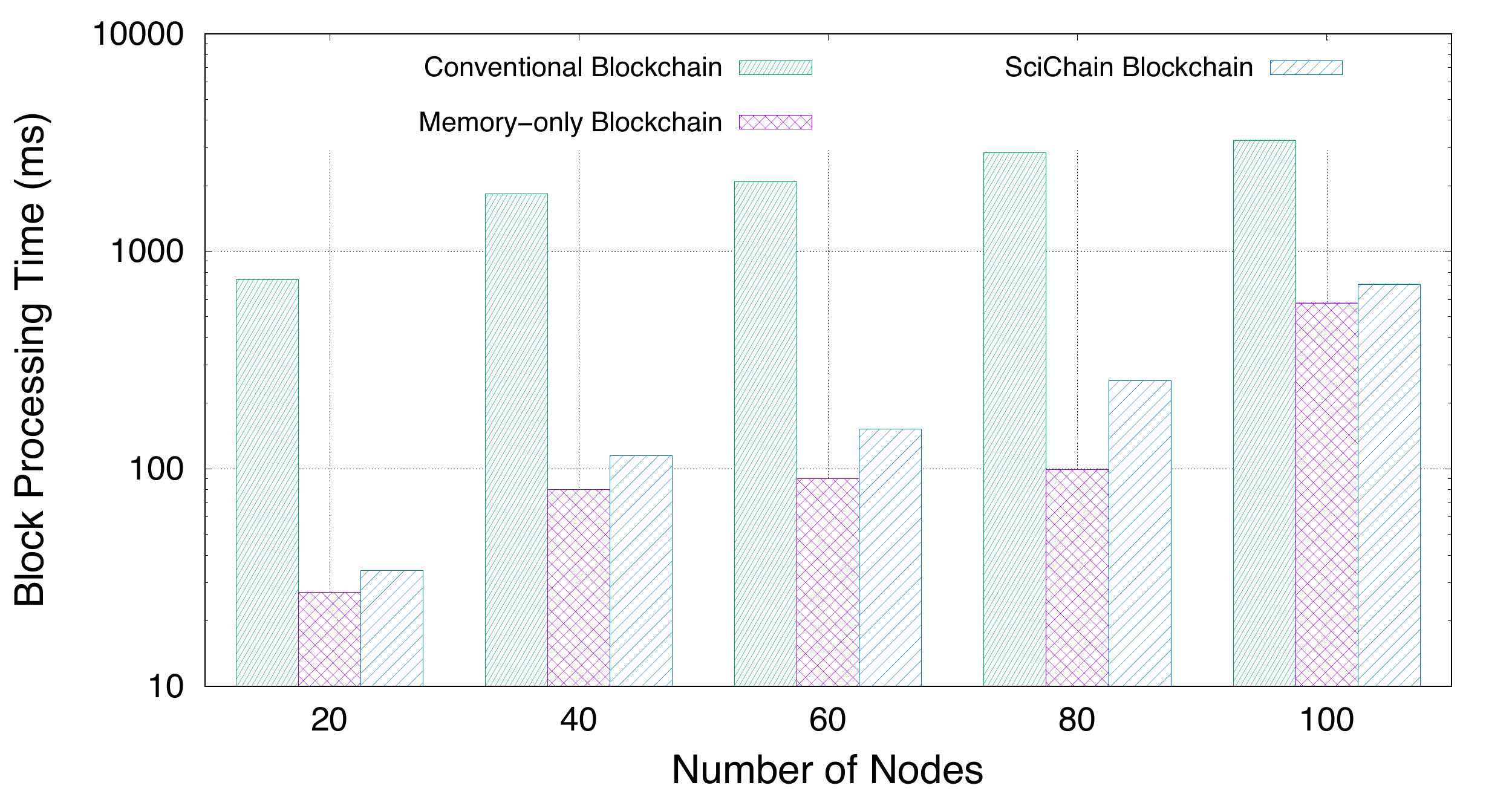}
    \caption{Scalability of three blockchain systems.
    }
    \label{fig:scaling}
    \figspace
\end{figure}

Figure~\ref{fig:scaling} reports the performance of the three systems on 20-, 40-, 60-, 80-, and 100-node scales.
The memory-only blockchain, as expected, achieves the highest performance (i.e., lowest processing time).
The proposed SciChain with shared storage does not exhibit significant slowdown than the upper bound;
for instance at 100-node scale, the comparison is 157 \textit{ms} vs. 311 \textit{ms},
at the same order of magnitude.
However, the conventional blockchain on 100 nodes appends a new block in 3,231 \textit{ms},
significantly slower than SciChain. 
Specifically, SciChain shows significant speedup in performance compared with existing systems:
$\frac{3,231}{311}$ = $10.4\times$.

\begin{figure*}[!th]
\centering
\subfigure{
    \includegraphics[width=140mm]{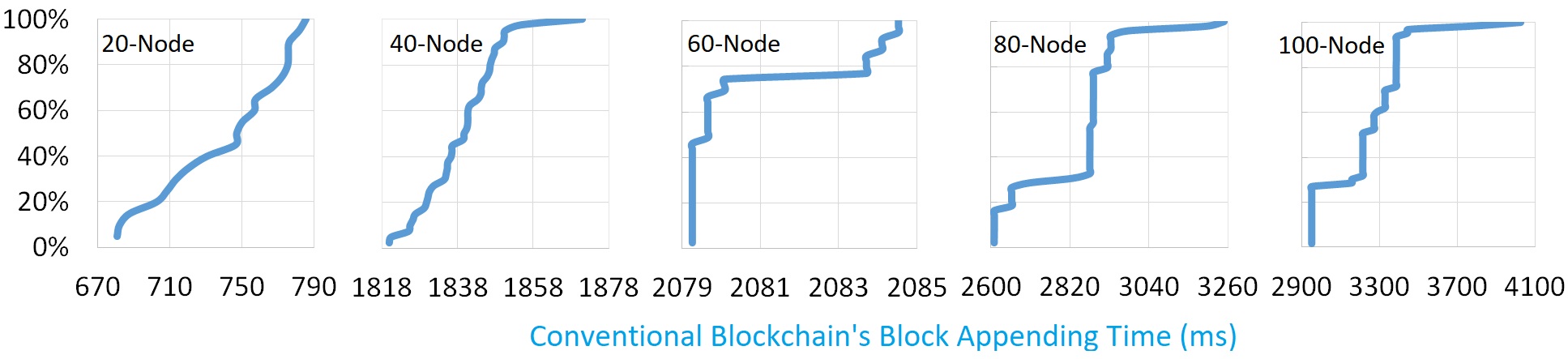}
    \label{fig:cdf_scale_ethernet}
}
\subfigure{
    \includegraphics[width=140mm]{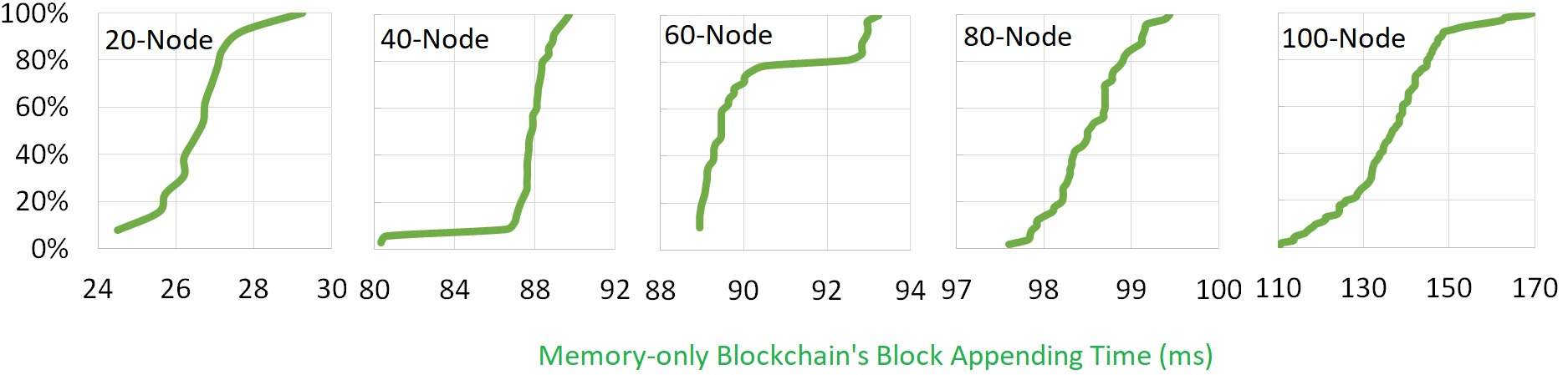}
    \label{fig:cdf_scale_rdma}
}
\subfigure{
    \includegraphics[width=140mm]{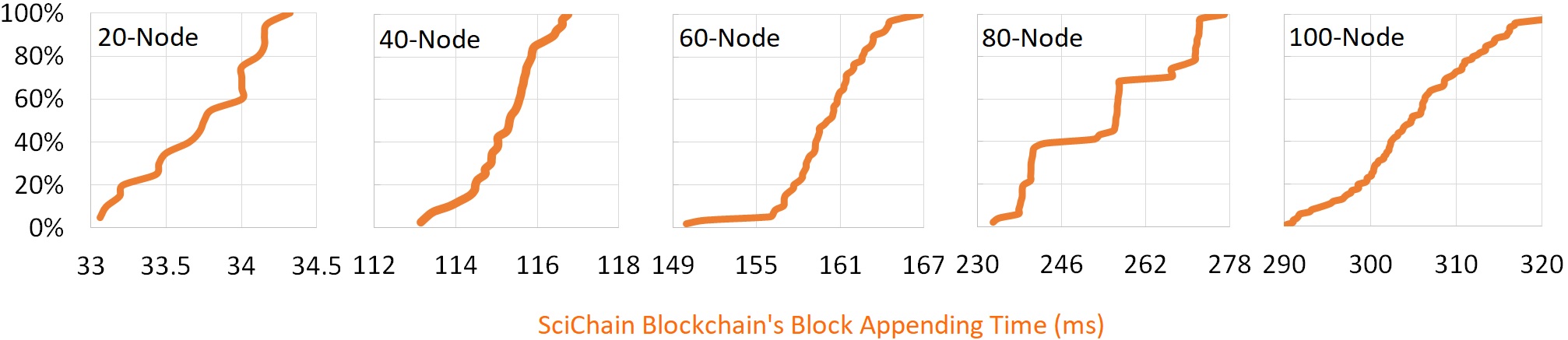}
    \label{fig:cdf_scale_imb}
}%%\figspace

\caption{CDFs of three blockchains' performance.
}
\label{fig:scalability}
\end{figure*}

In addition to reporting the average performance of three blockchain systems,
we draw the cumulative density function (CDF) of their performance in Figure~\ref{fig:scalability}.
The conventional blockchain shows long-tail issues at various scales (e.g., 60-node, 80-node, 100-node).
The memory-only blockchain also has long-tails problems multiple scales as well: 40-node and 60-node.
The proposed SciChain does not exhibit long tails and has a low and smooth variance at all scales.

\subsection{Large-Scale Applications}

We study a trace of the FusionFS~\cite{dzhao_tpds16} filesystem initially deployed to a 1,024-node cluster at the Argonne National Laboratory.
Four real-world scientific applications are being traced:
Plasma Physics, Turbulence, AstroPhysics, and Parallel BLAST~\cite{pblast_2003},
all of which run for five hours.
These applications cover a broad spectrum of I/O patterns exhibited by scientific applications~\cite{zzhang_hpdc13}.
We assign random numerical values to the I/O operations traced in FusionFS such that the SciChain prototype can take in the provenance in the same way as other blockchain systems.
We execute the workloads by feeding the transactions into the system prototype deployed on 1,024 nodes---the same scale when the applications were executed at the Argonne National Laboratory. 

\begin{figure}[!t]
    \centering
    \includegraphics[width=65mm]{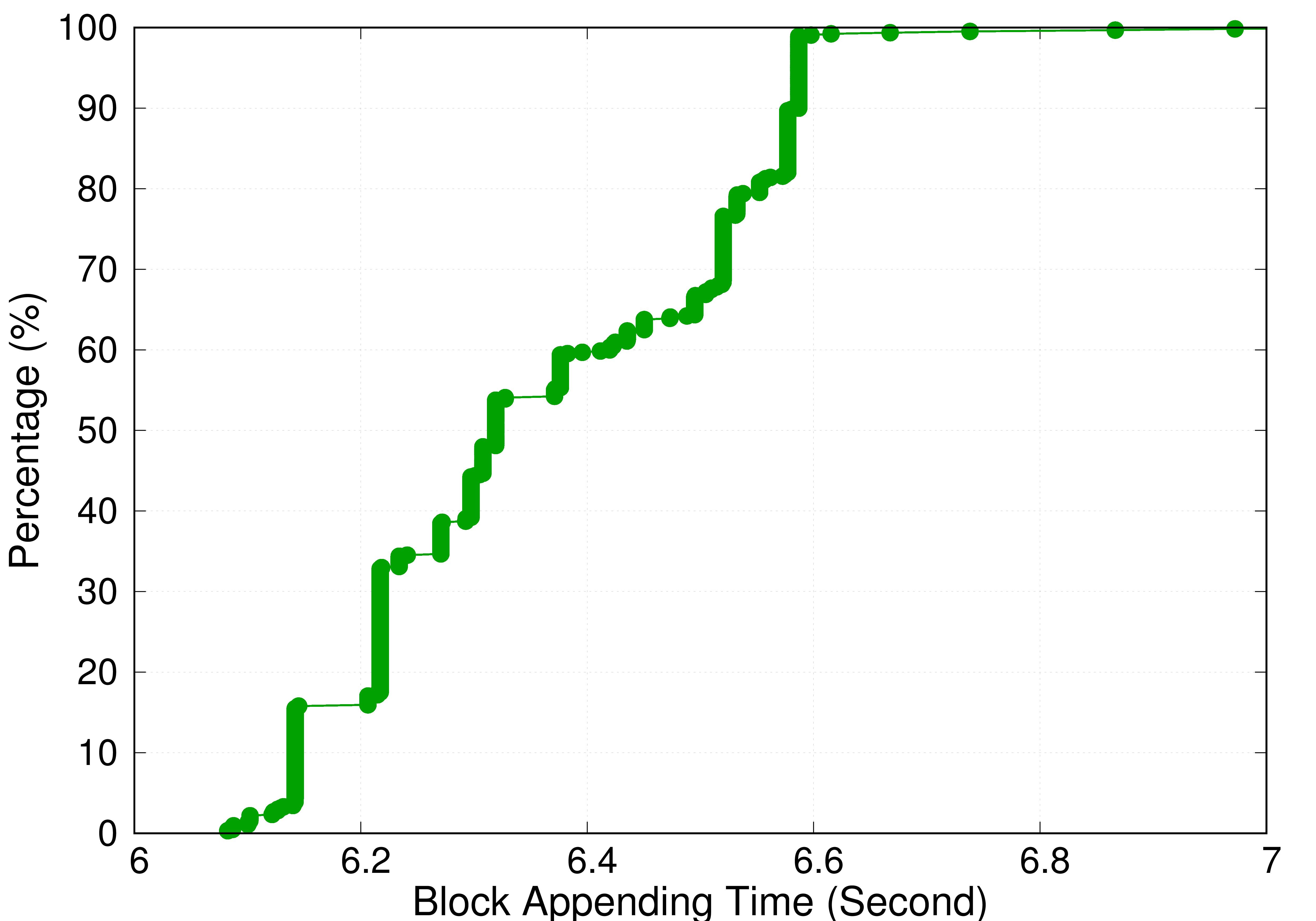}
    \caption{Latency distribution on 1,024 nodes. 
    }
    \label{fig:latency_1k_node}
    \figspace
\end{figure}

As shown in Figure~\ref{fig:latency_1k_node}, the 1,024-node cluster appends a new block in 6--7 seconds.
We do observe a few outliers (data points at 6.7 \textit{s} and beyond along the $X$-axis), 
although the quantity is tiny:
only four out of 1,024 nodes, less than 0.4\%.
The real difference is also insignificant,
about 0.4 seconds out of seven seconds.
It should be noted that, 
each block consists of more than 12 transactions.
Therefore, on 1,024 nodes, the provenance time incurred by SciChain is sub-second---a reasonable, if not negligible, overhead for large-scale, hours-long scientific applications in practice.

\section{Conclusion and Future Work}
\label{sec:conc}

This paper proposes a new blockchain consensus protocol, namely POST, to enable immutable and autonomous data provenance for scientific applications deployed to HPC systems.
POST is implemented in a prototype system called SciChain, the first HPC-blockchain system toward trustworthy data provenance in HPC.
The effectiveness and efficiency of POST are analyzed,
and experimentally demonstrated through both micro-benchmarks and real-world applications on up to 1,024 nodes.

Our future work is two-fold.
Firstly, the current POST protocol does not consider the existence of burst buffers available in many extreme-scale supercomputers.
We believe POST will be more efficient with the help of burst buffers as a near-local storage option.
Secondly, at this point, it is unclear how to migrate the data stored in one specific SciChain instance to another.
This is also a challenge in the blockchain community:
ensuring the atomicity of the transaction migration is far more complicated than it looks.

\section*{Acknowledgement}
% \textbf{Acknowledgement.} 
This work is in part supported by a Google Cloud Platform Research Award.
This work is also supported by the U.S. Department of Energy (DOE) under contract number DE-SC0020455.

{
\footnotesize
\bibliographystyle{IEEEtran}
\bibliography{ref_new}
}

\end{document}